\documentclass[12pt]{article}
\usepackage[dvips]{color}
\usepackage{amsmath}
\usepackage{graphicx}
\usepackage{subfigure}
\usepackage{epsfig}
\usepackage{amsmath}
\usepackage{cite}
\usepackage{color}
\usepackage{subfigure}

\begin{document}
\begin{center}
\Large{\bf Strong Cosmic Censorship in light of Weak Gravity Conjecture for Charged Black Holes}\\
\small \vspace{1cm} {\bf Jafar Sadeghi $^{\star}$\footnote {Email:~~~pouriya@ipm.ir}}, \quad
 {\bf Mohammad Reza Alipour $^{\star}$\footnote {Email:~~~mr.alipour@stu.umz.ac.ir}}, \quad
 {\bf Saeed Noori Gashti$^{\star,\ddag}$\footnote {Email:~~~saeed.noorigashti@stu.umz.ac.ir}}\\
\vspace{0.5cm}$^{\star}${Department of Physics, Faculty of Basic
Sciences,\\
University of Mazandaran
P. O. Box 47416-95447, Babolsar, Iran}\\
\vspace{0.5cm}$^{\ddag}${School of Physics, Damghan University, P. O. Box 3671641167, Damghan, Iran}\\

\end{center}
\begin{abstract}
In this paper, we investigate the strong cosmic censorship conjecture (SCC) for charged black holes in the de Sitter space by considering the weak gravity conjecture (WGC). Using analytical methods, we find that the SCC is preserved for dS-charged black holes with respect to some restriction $qQ\gg1$ and $r_+\geq Q$ with the help of the WGC condition viz $\frac{q}{m}\geq 1$ for scalar fields. Where q, m are the charge and mass of the scalar field, and $r_+$, Q determine the radius of the outer event horizon and the charge of the black hole, respectively. In that case, when the (WGC) is valid, SCC will definitely be satisfied for the dS-charged black holes. On the other hand, the SCC is violated when the WGC is not satisfied. Also, we examined the RN-dS charged black hole in the extremality state and found that SCC can be violated with the condition $\Lambda r_+^2=1$.
\end{abstract}
\tableofcontents
\section{Introduction}\label{into}
One of the studies with a long history in general relativity is the study of the collapse of small perturbations. We need more information on how these oscillations decay to understand better the gravity concept, use gravitational wave data, and study and investigate the valuable features of general relativity. One of the signs of the failure of determinism in general relativity can be the emergence of an interesting phenomenon known as Cauchy horizons that appear in the astrophysical solutions of Einstein's equations. These horizons are such that it is impossible to specify the history of the future of an observer that passes through such horizons using Einstein's equations and initial data. With these descriptions in the black holes' space-time background, it is an expected possibility that the perturbations of the outer region are infinitely amplified by a mechanism known as the blue shift. They lead to a singularity boundary beyond the Cauchy horizon in the interior of black holes, where field equations cease to make sense. The Penrose strong cosmic censorship (SCC) confirms such an expectation. Of course, another point is that astrophysical black holes are stable due to a special mechanism called the perturbation-damping mechanism, which is applied in the outer region. Therefore, considering whether SCC retains real hinges or not depends on the very subtle competition between the collapse of perturbations in the outer region and their amplification (blue shift) in the inner space-time of black holes. In general, the fate of Cauchy horizons is related to the collapse of small perturbations outside the event horizon. Hence, the validity of SCC is tied to the extent of external damps fluctuation. In connection with various structures and conditions, SCC and its satisfaction and violation have been investigated in various theories. The violation of this conjecture near the extremal region studied in the investigation of higher curvature gravity \cite{1}. Also, this conjecture has been challenged in investigating many charged black holes. In \cite{2,3}, this conjecture was checked for a charged AdS black hole. It was shown that for a specific interval for the parameter ($\beta$), this conjecture is satisfied and violated in other areas as well. The strong cosmic censorship conjecture has also been investigated in two dimensions. There have been interesting outcomes regarding the violation of this conjecture near the extremal region at specific points \cite{4}.
The study of this conjecture in the structure of three-dimensional black strings has also carried interesting results, which you can see \cite{5} for a deeper study.
Also, you can see \cite{6,7,8,9} for further study.
The effectiveness of mass-inflation systems, which are involved in the transformations of the inner Cauchy horizon associated with the space-time of black holes that are approximately flat, which is pathological in the estimation of SCC, into a series of hypersurfaces which is singular non-extendable. Those that are in an indivisible form are related to two different types of physical mechanisms\cite{b,c,d,e,f,g,h}.
First, the events in the exterior space-time regions of dynamic black holes formed viz the collapse of the remnant perturbation fields and second amplification of exponential blue shift related to the fields falling into the inside of black holes. We can manage these two introduced different systems through parameters such as ($g$) and ($k_{-}$).
It can be stated that the dimensionless physical ratio with the help of these two parameters can determine the fate of the inner Cauchy horizons inside such space-times of non-asymptotic flat black holes\cite{8,j,k},
\begin{equation*}\label{37}
\begin{split}
\beta\equiv\frac{g}{k_{-}}.
\end{split}
\end{equation*}
Of course,  a certain range of parameters of black holes, such as mass and charge, etc., as indicated in \cite{8,j,k},
\begin{equation*}\label{37}
\begin{split}
\beta>\frac{1}{2}.
\end{split}
\end{equation*}
So, space-time of the corresponding black holes can be physically expanded beyond their Cauchy horizon which includes a pathological fact and a sign of algebraic failure or a violation of the Penrose SCC in classical general relativity.
For the dynamics of Einstein's equations as well as the destiny of the observers, the explosive structure of the curvature that is related to ($\beta<1$) does not have per se much physical significance: it indicates two theorems, not the failure of the field equations mentioned in \cite{q} and of course not the destruction of macroscopic observers which is discussed in \cite{e}.
Therefore, the physical and mathematical formulation of the conjecture  of a SCC in such conditions leads to ignoring physical phenomena such as impulsive gravitational waves or the formation of shocks in relativistic fluids.
Due to the aforementioned reasons, the modern form of the SCC conjecture was introduced that requires a stronger constraint ($\beta<\frac{1}{2}$ ) and many works have been done to fit such constraints.
In \cite{8}, it was found that when there are \emph{neutral} scalar fields in the presence of the R-N dS black hole, it leads to the violation of SCC. But in continuation, Hod has shown in \cite{13} that the presence of \emph{charged} massive scalar fields near \emph{charged} black holes is inevitable. By considering \emph{charged} scalar fields near a \emph{charged} black hole and using WKB techniques, he has shown that the SCC will be meted. In this article, by using  an analytical method and  WKB approximation, we show  that \emph{charged} scalar fields play an essential role in satisfying the SCC in light of WGC for charged black holes. Therefore, we find that the numerical results presented in \cite{8} have no physical relevance to the question of the (in) validity of the SCC in \emph{charged} black-hole spacetimes. In particular, the SCC conjecture in the context of \emph{charged} black hole spacetimes must be tested in the presence of \emph{charged} matter fields, whereas the numerical results presented in \cite{8} are based only on the presence of \emph{neutral} scalar fields in the \emph{charged} black hole spacetime. Therefore, in this article, we are going  to study  different structure of this conjecture. According to the above description, we consider the general configuration of \emph{charged} black holes in the presence of massive \emph{charged} scalar fields. Then, using the weak gravity conjecture, we will prove that SCC is valid for specific values for all \emph{charged} black holes. We will use the weak gravity conjecture to prove a general relation with respect to SCC for all \emph{charged} black holes.
According to the above explanations, we organize the article in the following form.\\
In section 2, we will give basic explanations about the weak gravity conjecture and also the motivation to use it. In section 3, we  introduce charged black holes in dS space, and then we  show the quasinormal resonant frequency spectrum in section 4. We  check the conditions of compatibility and violation of (SCC) with respect to (WGC) for RN-dS charged black holes. Finally, we describe the results in section 5.

\section{Weak Gravity Conjecture}
As it is known in the literature, a new idea has been put forward as a swampland program to check theories coupled to gravity, the  consistency of quantum gravity and finally,  a proof for string theory. Recently, ones have done lots of work on this field \cite{bb,cc,dd,eee,fff,ggg,hhh,aaa,bbb,ccc,ddd}.
 Due to the special conditions of string theory and the fact that its testing and experimental investigations seem a bit difficult, this idea has been proposed to test and investigate various concepts of cosmology.
The swampland program is challenged from two sides.
From an up-bottom view for introducing principles and limitations to introduce conjectures, as well as mathematical formulations to examine cosmological concepts. A second look from the bottom-up in order to test each of these conjectures with various concepts of cosmology including inflation and matching with observable data, which is both a proof for this new idea and a proof for string theory.
So far, many conjectures have been proposed from this theory, and now, according to the structure and further investigations, new conjectures will be added to this program.
Some of these conjectures face challenges and as a result, corrections are made to the conjectures.
We face some limitations in quantum gravity (QG). At the point when gravity is considered at the quantum level, the hypothesis will be incompatible. Generally having a reliable quantum hypothesis of gravity isn't really straightforward and can in any case hold many surprises and can be interesting for physical science at low energies. The objective of the swampland program is to decide the limitations that an effective field theory(EFT) should fulfill to be viable with the consideration of ultraviolet completion(UV) in QG. They are called swampland limitations, and different suggestions are figured out as far as swampland conjectures(SC). The objective is to recognize these limitations, accumulate proof to demonstrate or refute them inside the structure of QG, give reasoning to make sense of them in a model-free manner, and comprehend their phenomenological suggestions for low-energy EFTs. Albeit the swampland idea isn't restricted to string theory on a fundamental level, SC are frequently examined by string theory backdrops. Without a doubt, the string theory gives an ideal structure to thorough quantitative testing of conjectures and works on how we might interpret potential compressions of string theory. Strangely, it has as of late been uncovered that a large number of these conjectures are to be sure related, recommending that they may essentially be various countenances of some yet-to-be-found crucial standard of QG. As far as possible have significant ramifications for cosmology and particle physics. They can give new core values to building conjectures past the standard models in high-energy physics. They may likewise prompt $UV/IR$ blending, which breaks the assumption for scale detachment and possibly gives new bits of knowledge into the regular issues seen in our universe. Consequently, the presence of swampland is extraordinary information for phenomenology. For a total rundown of references connected with swampland that might be valuable, we allude in \cite{bb} the swampland program (SP) has likewise been surveyed and presented. The shortfall of global symmetry (GS) and the completeness of charge spectra are at the center of the SP. Nonetheless, they need phenomenological suggestions except if we can restrict the global symmetries \cite{cc,dd} and whether there is any limitations point on the mass of charged states. In any case, they just bound the complete hypothesis but not the low-energy EFTs. Specifically, it is phenomenologically important whether all charged particles can be really super heavy and even compare to black holes(BHs), or whether there is some thought of completeness of the range that gets by at low energies. A large portion of the SCs examined address exactly these inquiries. They want to profoundly explore these assertions and measure them so we can draw nearer to the recuperation of a few global symmetries. For instance, we can deduce recuperate a global symmetry (GS) $U(1)$ by sending the gauge coupling(GC) to nothing, which ought not to be permitted in $QG$. Attempting to comprehend string theory for the study of this issue,  it might turn out that if one somehow managed to try to do such work, can give data about the imperatives that an EFT can fulfill to be viable with QG. Likewise, WGC forbids this cycle by flagging the presence of new charged states that denies the depiction of the EFTs. Thusly, it gives an upper bound on the mass of these charged states. The WGC comprises of some parts: the electric and the magnetic
electric-WGC: As indicated by a quantum hypothesis, we have the following condition \cite{bb,cc,dd,eee,fff,ggg,hhh,aaa,bbb,ccc,ddd},
\begin{equation}\label{eq177}
\begin{split}
\frac{Q}{m}\geq\frac{\mathcal{Q}}{M}|_{ext}=\mathcal{O}(1),
\end{split}
\end{equation}
and
\begin{equation}\label{eq188}
\begin{split}
Q=g q,
\end{split}
\end{equation}
where, $g$ and $q$ are the gauge coupling and the quantized charge.
The electric-WGC needs the presence of an electrically charged condition of a higher charge-to-mass proportion than extremal BH in that hypothesis, which is regularly a variable of the order one.
One more understanding of this conjecture  is that the limitations region shows that scale force determines stronger than the  gravity on this mode — so subsequently is called WGC. This is an identical equation since it expects that electromagnetic force is stronger gravitational force \cite{bb,cc,dd,eee,fff,ggg,hhh,aaa,bbb,ccc,ddd},
\begin{equation}\label{eq200}
F_{Grav}\leq F_{EM}
\end{equation}
It implies that the charge is more prominent than the mass, so we get a similar condition as above. This is as of now false within the sight of massless scalar fields.
The motivations of $WGC$ are twofold. To begin with, it makes a $QG$ boundary to reestablish the GS of $U(1)$ by sending $g\rightarrow 0$. If a GC goes to zero as indicated by $WGC$, this conducts new light particles and the cutoff the hypothesis arrives at nothing and nullifies the EFT. Because of the littleness of the scale coupling, it relies upon how much energy the interaction with which you need to portray the viable EFT. The smallness of the cycle energy leads to the smallness of the scale coupling. On the other hand, if you need to keep the EFT substantial up to an extremely high cut-off, the GC can't be excessively small. This is an illustration of swampland limitations that becomes more grounded for higher energies. Obviously, a hypothesis with disappearing measure coupling i.e., GS is incompatible because the cutoff of the viable EFT is likewise zero. One more fundamental inspiration for $WGC$ is that a kinematic prerequisite permits extremality $BH$ to have decomposed. Charged BHs should fulfill an extremality breaking point to stay away from the presence of exposed singularities, as shown by the weak cosmic censorship $(WCC)$. For a given charge $\mathcal{Q}$, this super bound shows that the this super bound shows that the mass $M$ of the BHs should be more noteworthy than the charge \cite{bb,cc,dd,eee,fff,ggg,hhh,aaa,bbb,ccc,ddd},
\begin{equation}\label{eq288}
M \geq \mathcal{Q}
\end{equation}
For the BHs to have a regular horizon. Here, we set the extremal factor $\mathcal{O}(1)$ to one for simplicity. The primary condition for starting the decay to the small black hole and particle is the existence of the extremal BHs ($M=\mathcal{Q}$). So, one can consider the decay of an extremal BHs which one of the rot items has a charge more modest than its mass as far as possible, so $M_{1}\geq \mathcal{Q}_{1}$. Then the rot item can never again have a charge more modest than the mass, that is $m_{2}\leq Q_{2}$. It is just a kinematic necessity. Since the second rot item violates the WCC, it can't be a BH, so it should be a particle. The above kinematic necessity can be acquired by applying preservation of mass/energy and protection of charge as follows. The initial mass of the  BH should be more prominent than the amount of the mass of the rot items $M_{i}$ and the charge of the initial BH.

\section{Charged Black Holes in dS Space }\label{s2}
The metric of charged black hole in spherical symmetric space is defined as follows,
\begin{equation}\label{eq1}
dS^2=f(r)dt^2-f^{-1}(r)dr^2-r^2 d\Omega^2, \qquad     d\Omega^2=(d\theta^2+\sin^2(\theta)d\varphi^2).
\end{equation}
Here, we consider $f(r)=H(M,Q)-\frac{\Lambda r^2}{3}$ in general; where Q, M, $\Lambda>0$ are electric charge, the mass of the  black hole and the cosmological constant respectively. In this case, we can obtain its event horizons as follows,
\begin{equation}\label{eq2}
f(r_\star)=0 \qquad \rightarrow   \qquad    \star \in (-,+,...,c).
\end{equation}
Considering the metric in general terms, we have different event horizons, where ($r_-$) is the Cauchy horizon, ($r_+$) is the outer event horizon, and ($r_c$) is the cosmological horizons.
Using Klein-Gordon's differential equation, we can determine the dynamics of a massive charged particle near a charged black hole \cite{10,11,12,13},
\begin{equation}\label{eq3}
\frac{1}{\sqrt{-g}}\partial_{\mu}(g^{\mu\nu}\sqrt{-g}\partial_{\nu}\Phi)-2iq g^{\mu\nu}A_{\mu}\partial_{\nu}\Phi-q^2 g^{\mu\nu}A_{\mu}A_{\nu}\Phi-m^2\Phi=0,
\end{equation}
where $m$ and $q$ are the mass and charge of the particle, respectively also,  $A_\mu=\left(\frac{Q}{r},0,0,0\right)$. We can define the scalar field $\Phi$ according to relation \eqref{eq3} as follows \cite{15},
\begin{equation}\label{eq4}
\Phi(t, r, \theta, \phi) =\sum_{m} \sum_{\ell}e^{-i\omega t} Y_{\ell m}(\theta, \varphi) \Phi(r).
\end{equation}
The integer parameters $\ell$ and $m$  are the spherical and the azimuthal harmonic indices of the resonant eigenmodes which characterize the charged massive scalar fields in the charged black-hole spacetime.
By putting Eq.\eqref{eq4} in Eq.\eqref{eq3} and using $dx=\frac{dr}{f(r)}$, we get the Schrödinger-like differential equation ,
\begin{equation}\label{eq5}
\frac{d^2 \phi(r)}{dx^2}+V(r) \phi(r)=0.
\end{equation}
The effective radial potential due to a massive charged particle near a charged black hole is defined as \cite{8},
\begin{equation}\label{eq6}
V(r)=\frac{qm}{r^2}\left[\frac{q}{m} \alpha(r) -\frac{m}{q}\beta(r)\right],
\end{equation}
where
\begin{equation}\label{eq7}
\alpha(r)=Q^2\left(1-\frac{\omega r}{qQ}\right)^2, \qquad  \beta(r)=r^2 f(r) H(r), \qquad H(r)=\left(\frac{\ell(\ell+1)}{m^2 r^2}+\frac{f^{\prime}(r)}{m^2 r}+1\right).
\end{equation}
Also, we can consider the boundary conditions for the special radial function near the outer event horizon as an incoming wave and at the largest event horizon as an outgoing wave \cite{13,14}:
\begin{eqnarray}\label{eq8}
\phi(x) \sim \left\{
               \begin{array}{ll}
                 e^{-i (\omega-\frac{qQ}{r_+})x}, & \hbox{for \hspace{0.2cm} $r \rightarrow r_+ \hspace{0.1cm}(x \rightarrow -\infty)$;} \\
                 e^{-i (\omega-\frac{qQ}{r_c})x}, & \hbox{for  \hspace{0.2cm} $r \rightarrow r_c \hspace{0.1cm}(x \rightarrow \infty)$.}
               \end{array}
             \right.
\end{eqnarray}
According to the above boundary conditions, we can obtain the discrete spectrum of $\omega$, defined as the resonance frequency of the imaginary quasi-normal state.
\section{The Quasinormal Resonant Frequency Spectrum }\label{s3}
In this section, we need to obtain the imaginary part of the resonance frequency to investigate the linear dynamics of a massive charged particle near a general charged black hole.
Also, we need to do this in a dimensionless regime to do this analytically. Since, the $\frac{q^2}{\hbar}\simeq \frac{1}{137}$ relationship exists in our universe, we can consider it for black holes, even slightly charged, and get $qQ \gg 1$. In addition, the mechanism of the Schwinger-type pair-production in space-time of charged black hole creates a limit to the black hole electric field with the $\frac{Q}{r_+^2}\ll \frac{m^2}{q}$ relationship \cite{17,18,19,20}. Therefore, according to the above statement, we can consider SCC and define our constraint regime following ansans,
\begin{equation}\label{eq9}
m^2 r_+^2 \gg \ell(\ell+1)  \qquad  and \qquad    m^2 r_+^2\gg 2 k_+ r_+,
\end{equation}
where $k_+=f^{\prime}(r_+)/2$ is the gravitational acceleration of the black hole at the outer event horizon. In this area, we try to obtain the imaginary part of the resonance frequency in the background of the general charged black hole near the event horizon. Now, we use radial potential \eqref{eq6} to determine the linear dynamics of the massive charged particle near the event horizon of the black hole. We can consider this potential in region \eqref{eq9} as an effective potential and obtain the quasinormal resonance modes analytically using standard WKB techniques \cite{21,22}. In this region, we consider the maximum effective potential near the event horizon of the black hole at point $r=r_0$. In the following, we use the relationship \eqref{eq6}, \eqref{eq7}, and $V^{\prime}(r_0)=0$ to obtain the point where the effective potential is maximum as follows,
\begin{equation}\label{eq10}
r_0=\frac{q^2Q^2}{qQ\omega-m^2r_+^2 k_+}
\end{equation}
According to the Schrödinger-like differential equation \eqref{eq5} and \cite{21,22,23}, we use the $WKB$ method to obtain the quasinormal mode frequencies through the following,
\begin{equation}\label{eq11}
iK-(n+\frac{1}{2})-\Lambda(n)=\Omega(n)
\end{equation}
where
\begin{equation}\label{eq12}
\begin{split}
&K=\frac{V_0}{\sqrt{2 V_0^{(2)}}} \\
&\Lambda(n)=\frac{1}{\sqrt{2 V_0^{(2)}}}\left[\frac{\left(\alpha ^2+\frac{1}{4}\right)}{8 }\frac{V_0^{(4)}}{V_0^{(2)}}-\frac{\left(60 \alpha ^2+7\right)}{288}  \left(\frac{V_0^{(3)}}{V_0^{(2)}}\right)^2\right]\\
&\Omega(n)=\frac{n+\frac{1}{2}}{2 V_0^{(2)}}\left[\frac{5 \left(188 \alpha ^2+77\right)}{6912} \left(\frac{V_0^{(3)}}{V_0^{(2)}}\right)^4
-\frac{\left(100 \alpha ^2+51\right) }{384 } \frac{\left(V_0^{(3)}\right)^2 V_0^{(4)}}{\left(V_0^{(2)}\right)^3}\right]\\
&+\frac{n+\frac{1}{2}}{2 V_0^{(2)}}\left[ \frac{\left(68 \alpha ^2+67\right) }{2304}\left(\frac{V_0^{(4)}}{V_0^{(2)}}\right)^2 +\frac{\left(28 \alpha ^2+19\right) }{288 }\frac{\left(V_0^{(3)} V_0^{(5)}\right)}{\left(V_0^{(2)}\right)^2}-\frac{\left(4 \alpha ^2+5\right) }{288}\frac{V_0^{(6)}}{ V_0^{(2)}} \right]
\end{split}
\end{equation}
Here, $V_0^{(k)}\equiv\frac{d^kV}{dx^k}|_{r=r_0}$ is the spatial derivative of the effective potential of equation \eqref{eq6}, and its scattering peak is evaluated at the point $r=r_0$.
Using relations \eqref{eq6}, \eqref{eq7}, \eqref{eq10} and \eqref{eq12}, we will have the following relation in the region of \eqref{eq9},
\begin{equation}\label{eq13}
\begin{split}
&K\simeq\frac{k_+^2 m^4 r_+^4q Q }{2f_0 \left(k_+ m^2 r_+^2-q Q \omega\right)^2}\\
&\Lambda(n)\simeq\frac{k_+^2 m^4 \left[17-60 \left(n+\frac{1}{2}\right)^2\right] r_+^4+2 k_+ m^2 \left[36 \left(n+\frac{1}{2}\right)^2-7\right] q Q r_+^2 \omega}{16 q Q \left(q Q \omega-3 k_+ m^2 r_+^2\right)^2}\times f_0\\
&\mathcal{A}=15 k_+^4 m^8 \left[148 (n+\frac{1}{2})^2-41\right] r_+^8+12 k_+^3 m^6 \left[121-420 (n+\frac{1}{2})^2\right] q Q r_+^6 \omega\\
&\mathcal{B}=64 q^5 Q^5 \left(k_+ m^2 r_+^2-q Q \omega\right)^4\\
&\Omega(n)\simeq-(n+\frac{1}{2}) q^3 Q^3 f_0^2\times \frac{\mathcal{A}}{\mathcal{B}}
\end{split}
\end{equation}
where $f_0=f(r_0)$. In the next step, to determine the study of SCC, we need to obtain the minimum value of the fundamental imaginary resonance mode of the system. For this purpose, using equations \eqref{eq11} and \eqref{eq13}, we can calculate the $Im(\omega_0)$,
\begin{equation}\label{eq14}
\begin{split}
&\omega\simeq \frac{qQ}{r_+}-\frac{2k_+ m^2 r_+^2}{qQ}\left[1-\frac{14400}{11644}\left(\frac{(n+1/2)f_0}{qQ}\right)^4\right]\\
&-i\left\{4f_0 k_+(n+\frac{1}{2})\frac{m^2 r_+^2}{q^2Q^2} \left[1-\frac{34 qQ f_0^4}{11664}\right]+\mathcal{O}(f_0^2)\right\}
\end{split}
\end{equation}
Since we consider $r_0$ near the event horizon ($r_+$), we have $f_0\ll 1$. For investigation the SCC, it is necessary to find the minimum value of the resonance mode and evaluate its ratio to the surface gravity of the event horizon,
\begin{equation}\label{eq15}
\beta=\frac{-Im(\omega_0)}{k_+}\simeq2f_0 \frac{m^2 r_+^2}{q^2Q^2} \left[1-\frac{34 qQ f_0^4}{11664}\right].
\end{equation}
Since it is $f_0\ll 1$, it is sufficient to have the conditions $q^2Q^2>m^2r_+^2$ in the relation above concepts so that $\frac{-Im(\omega_0)}{k_+}<\frac{1}{2}$ is established.
Therefore, we have the following condition for the study of SCC,
\begin{equation}\label{eq16}
\frac{q}{m}\geq \frac{r_+}{Q}.
\end{equation}
from equation \eqref{eq16} determine that when $r_+\geq Q$, we have the weak gravity conjecture condition. We  know that $k_->k_+$, so the relationship of \eqref{eq15} and \eqref{eq16} is also established for $\beta=\frac{-Im(\omega_0)}{k_-}<\frac{1}{2}$. Also, according to relation \eqref{eq15}, when $qQ<2\sqrt{f_0} mr_+$, SCC can be violated. Since $qQ\gg 1$ and $f_0\ll 1$, the mass of the scalar field and the radius of the event horizon must be very massive and very large respectively.
In the following, we obtain the extremality state of the RN-dS black hole. We will have the following relation for the RN-dS black hole with respect to equation\eqref{eq1},
\begin{equation}\label{eq17}
f(r)=1-\frac{2M}{r}+\frac{Q^2}{r^2}-\frac{\Lambda r^2}{3}.
\end{equation}
When $k_+=k_-=0$, we can obtain the  black hole extremality state,
\begin{equation}\label{eq18}
Q_{exe}=r_+\sqrt{1-\Lambda r_+^2}, \qquad    M_{exe}=r_+(1-\frac{2}{3}\Lambda r_+^2).
\end{equation}
We substitute Eq.\eqref{eq18} in  Eq.\eqref{eq15} to obtain $\beta$ in the extremality state of the RN black hole,
\begin{equation}\label{eq19}
\beta \simeq2f_0 \frac{m^2 }{q^2(1-\Lambda r_+^2)} \left[1-\frac{34 q f_0^4 \sqrt{1-\Lambda r_+^2} r_+}{11664}\right]
\end{equation}
According the above relationship, when the condition $\frac{q}{m}\geq\frac{1}{1-\Lambda r_+^2}$ is satisfied, the SCC will definitely be preserved, and since $\Lambda r_+^2<1$, the weak gravity conjecture will also be satisfied. On the other hand, when $\Lambda r_+^2\ll1$, we will have the $SCC$ condition in light of the $WGC$,
\begin{equation}\label{eq20}
\frac{q}{m}\geq 1+\Lambda r_+^2,
\end{equation}
from the above relation WGC is clearly obtained. In relation \eqref{eq19} when $\Lambda r_+^2=1$, we have $\beta \rightarrow \infty$ and the SCC is violated. Also, these result and conditions are completely compatible with \cite{24,25}.

\section{Discussion and Result}\label{s51}
One of the indications of the failure of determinism GR can be the rise of a fascinating peculiarity known as the Cauchy horizon that shows up in the astrophysical solutions of Einstein's equations. These horizons are such that it is difficult to indicate the history of the future of an observer that passes come of such horizons utilizing Einstein's conditions and initial information. With these descriptions in the black holes' background space-time, it is a predicted possibility that the perturbations of the external area are infinitely enhanced by a system known as the blue shift. They lead to a singularity beyond the Cauchy horizon the inside of BHs, where field conditions fail to seem good. The Penrose cosmic censorship conjecture (SCC) affirms such an assumption. Obviously, another point is that astrophysical BHs are stable because of an exceptional component called the perturbation-damping system, which is applied in the outer region.
Also, the SCC resolves the issue of the idea of the singularities tracked down in many answers to Einstein's gravitational field equations: Are such singularities conventionally described by unbounded curvature? Is the presence of a Cauchy horizon, an unsteady characteristic element of answers of Einstein's equations? Recently researchers,  remarking on the historical backdrop of the SCC conjecture, overview a portion of the headway made in research coordinated either toward satisfying SCC or toward revealing a portion of its shortcomings. They specifically around model adaptations of SCC which have been demonstrated for constrained groups of spacetimes viz the Gowdy spacetimes and the role played by the conventional presence of Asymptotically speed term dominated conduct in these answers. Also additionally note some work on spacetimes containing weak null singularities, and their importance for the SCC \cite{24,25,1001}.
SCC conjecture has been one of the main acts of pure confidence with regard to GR, confirming the deterministic idea of the related field relations.
However, it holds well for asymptotically level spacetimes, an expected disappointment of the SCC conjecture could emerge for spacetimes acquiring Cauchy horizon alongside a positive cosmological constant viz its potential failure about this issue.
Researchers have unequivocally exhibited that infringement of the restriction SCC turns out as expected within the sight of a Maxwell field even with the presence of higher spacetime aspects. Specifically, for higher dimensions of the  RN black holes, the infringement of SCC is at a bigger scope compared with the 4D case, for specific of the cosmological constant.
Then again, for a brane world BH, the impact of an additional dimension is to make the infringement of cosmic censorship weaker.
For rotating BHs, intriguingly, the SCC is constantly holding even in the presence of higher dimensions.
A comparable situation is likewise noticed for rotating BHs on the brane \cite{1001}. In this paper, we investigated dynamically formed charged black holes.
Also, to satisfy the SCC, the inner Cauchy horizons of the black hole must be unstable.
Here, to check the SCC, it is necessary to get two $-Im (\omega_0)$ and $k_-$ parameters to demonstrate the decay rate of the remaining perturbation fields in the outer regions of the black hole and the blue-shift growth rate of the in-falling fields of the black hole, respectively. Therefore, if $\beta=\frac{-Im (\omega_0)}{k_-}<1/2$, SCC will be maintained. We found that for the dS charged black hole with respect to $r_+\geq Q$ in light of the WGC, viz $q/m\geq1$, SCC will definitely be satisfied. We also found that there will be a possibility of violation of SCC for the massive scalar field as well as when the radius of the event horizon of the charged black hole is very large. We also found SCC will be violated in the extremality state for the charged RN-dS black hole when $\Lambda r_+^2=1$ which is also mentioned in \cite{24,25}. Also, these results and conditions are completely compatible with \cite{24,25}. On the other hand, in \cite{8,26}, when the scalar field is uncharged, the SCC is violated, which is consistent with  \eqref{eq15} in this paper. Because can be obtained $\beta>1/2$ if assume the charge of the scalar field is zero viz $q = 0$.
The above study also raises some questions as follows.\\
Is the relationship researched in this article also valid for black holes in higher dimensions?
Do other black holes in different frames satisfy the SCC and WGC simultaneously?
Is it possible to consider the SCC relation with WGC monitoring for all black holes?
Is it may such a structure also be established for black holes on the brane?
We leave these questions for future work.

\section{Acknowledgments}
The authors would like to thank the referee for the fruitful comments to improve the introduction section.


\begin{thebibliography}{11}
\bibitem{1}
A. K. Mishra and S. Chakraborty, strong cosmic censorship conjecture in higher curvature gravity, Phys. Rev. D 101, 064041 (2020).
\bibitem{2}
C. Singha, S. Chakraborty and  N. Dadhich, Strong cosmic censorship conjecture for a charged BTZ black hole, J. High Energ. Phys. 2022, 28 (2022).
\bibitem{3}
C. Singha and  N. Dadhich, Strong cosmic censorship conjecture for a charged AdS black hole, arXiv:2209.10153 (2022).
\bibitem{4}
U. Moitra, Strong cosmic censorship in two dimensions, Phys. Rev. D 103, L081502 (2021).
\bibitem{5}
J. Ho, W. Kim, W. Kim and B. H. Lee, Investigations of strong cosmic censorship in 3-dimensional black strings, J. High Energ. Phys. 2022, 18 (2022).
\bibitem{6}
C. Y. Shaoa, L. J. Xina, W. Zhangb, C. G. Shaoa, Strong cosmic censorship for a charged black hole surrounded by quintessence, Physics Letters B  835, 137512 (2022).
\bibitem{7}
M. Casals and C. I. S. Marinho, Glimpses of violation of strong cosmic censorship in rotating black holes, Phys. Rev. D 106, 044060 (2022).
\bibitem{8}
V. Cardoso, J. L. Costa, K. Destounis, P. Hintz and A. Jansen, Quasinormal Modes and Strong Cosmic Censorship, Phys. Rev. Lett. 120, 031103 (2018).
\bibitem{9}
R. A. Konoplya and A. Zhidenko, How general is the strong cosmic censorship bound for quasinormal modes?, Journal of Cosmology and Astroparticle Physics 2022, (2022).



\bibitem{b}
P.R. Brady, I.G. Moss and R.C. Myers, Cosmic Censorship: As Strong As Ever, Phys. Rev. Lett. 80, 3432 (1998).
\bibitem{c}
W.A. Hiscock, Evolution of the interior of a charged black hole, Phys. Lett. A 83, 110 (1981).
\bibitem{d}
E. Poisson and W. Israel, Internal structure of black holes, Phys. Rev. D 41, 1796 (1990).
\bibitem{e}
A. Ori, Inner structure of a charged black hole: An exact mass-inflation solution, Phys. Rev. Lett. 67, 789 (1991).
\bibitem{f}
M. L. Gnedin and N. Y. Gnedin, Destruction of the Cauchy horizon in the Reissner-Nordstrom black hole, Class. Quantum Gravity 10, 1083 (1993).
\bibitem{g}
P.R. Brady and J.D. Smith, Black Hole Singularities: A Numerical Approach, Phys. Rev. Lett. 75, 1256 (1995).
\bibitem{h}
S. Hod and T. Piran, Mass Inflation in Dynamical Gravitational Collapse of a Charged Scalar Field, Phys. Rev. Lett. 81, 1554 (1998).
\bibitem{j}
M. Dafermos, Black Holes Without Spacelike Singularities, Commun. Math. Phys. 332, 729 (2014).
\bibitem{k}
P. Hintz and A. Vasy, Analysis of linear waves near the Cauchy horizon of cosmological black holes, J. Math. Phys. 58, 081509 (2017).
\bibitem{q}
S. Klainerman, I. Rodnianski, and J. Szeftel, The bounded $L^2$ curvature conjecture, Invent. Math. 202, 91 (2015).


\bibitem{bb}
E. Palti, The Swampland: Introduction and Review, Fortsch. Phys., 67, 1900037 (2019).
\bibitem{cc}
S. Fichet and P. Saraswat, Approximate Symmetries and Gravity, JHEP. 01, 088 (2020).
\bibitem{dd}
T. Daus, A. Hebecker, S. Leonhardt and J. March-Russell, Towards a Swampland Global
Symmetry conjcture using weak gravity, Nucl. Phys. B. 960, 115167 (2020).
\bibitem{eee}
M. van Beest, J. Calderón-Infante, D. Mirfendereski and I. Valenzuela,  Lectures on the swampland program in string compactifications. Physics Reports 989, 1-50 (2022).
\bibitem{fff}
E. Palti, A Brief Introduction to the Weak Gravity Conjecture, LHEP 2020,  176 (2020).
\bibitem{ggg}
C. Vafa, The String landscape and the swampland, hep-th/0509212.
\bibitem{hhh}
N. Arkani-Hamed, L. Motl, A. Nicolis and C. Vafa, The String landscape, black holes
and gravity as the weakest force, JHEP 06, 060 (2007).
\bibitem{aaa}
J. Sadeghi, M. Shokri, S. N. Gashti and M. R. Alipour, RPS thermodynamics of Taub–NUT AdS black holes in the presence of central charge and the weak gravity conjecture. Gen Relativ Gravit 54, 129 (2022).
\bibitem{bbb}
J. Sadeghi, M. Shokri, M. R. Alipour and S. N. Gashti, Weak gravity conjecture from conformal field theory: a challenge from hyperscaling violating and Kerr-Newman-AdS black holes, Chinese Physics C 47, 015103 (2023).
\bibitem{ccc}
J. Sadeghi, S. N. Gashti and E N. Mezerji, The investigation of universal relation between corrections to entropy and extremality bounds with verification WGC, Physics of the Dark Universe 30, 100626 (2020).
\bibitem{ddd}
J. Sadeghi, B. Pourhassan, S. N. Gashti and S. Upadhyay, Weak Gravity Conjecture, Black Branes and Violations of Universal Thermodynamic Relation, Annals of Physics 447, 169168 (2022).







\bibitem{10}
R. A. Konoplya and A. Zhidenko, Massive charged scalar field in the Kerr-Newman background: Quasinormal modes, late-time tails and stability, Phys. Rev. D 88, 024054 (2013).
\bibitem{11}
S. Hod, Mode-Coupling in Rotating Gravitational Collapse of a Scalar Field, Phys. Rev. D 61, 024033 (2000).
\bibitem{12}
E. N. Mezerji, J. Sadeghi, The correlation of WGC and hydrodynamics bound with R4 correction in the charged AdSd+ 2 black brane, Nuclear Physics B 981, 115858 (2022).
\bibitem{13}
S. Hod, Strong cosmic censorship in charged black-hole spacetimes: As strong as ever, Nuclear PhysicsB 941, 636 (2019).
\bibitem{14}
R.A. Konoplya, A. Zhidenko, Charged scalar field instability between the event and cosmological horizons, Phys. Rev. D 90, 064048 (2014).
\bibitem{15}
R. A. Konoplya and A. Zhidenko, Quasinormal modes of black holes: From astrophysics to string theory, Rev. Mod. Phys. 83, 793 (2011).
\bibitem{17}
J. Schwinger, On Gauge Invariance and Vacuum Polarization, Phys. Rev. 82,  664 (1951).
\bibitem{18}
B. Carter, Charge and Particle Conservation in Black-Hole Decay, Phys. Rev. Lett. 33,  558 (1974).
\bibitem{19}
S. Hod, Best approximation to a reversible process in black-hole physics and the area spectrum of spherical black holes, Phys. Rev. D 59,  024014 (1999).
\bibitem{20}
W. T. Zaumen, Upper bound on the electric charge of a black hole, Nature 247, 531 (1974).
\bibitem{21}
S. Iyer and C.M. Will, Black-hole normal modes: A WKB approach. I. Foundations and application of a higher-order WKB analysis of potential-barrier scattering, Phys. Rev. D 35, 3621 (1987).
\bibitem{22}
S. Iyer, Black-hole normal modes: A WKB approach. II. Schwarzschild black holes, Phys. Rev. D 35, 3632 (1987).
\bibitem{23}
B. F. Schutz and C. M. Will, Black hole normal modes - A semianalytic approach, Astrophys. J. 291,  L33 (1985).
\bibitem{24}
O. J. C. Dias, H. S. Reall, and J. E. Santos, Strong cosmic censorship for charged de Sitter black holes with a charged scalar field, Class. Quantum Grav. 36,  045005 (2019).
\bibitem{25}
O. J. C. Dias, H. S. Reall, and J. E. Santos, Strong cosmic censorship: taking the rough with
the smooth, J. High Energ. Phys. 2018, 1 (2018).
\bibitem{26}
V. Cardoso, J. L. Costa, K. Destounis, P. Hintz and A. Jansen, Strong cosmic censorship in charged black-hole spacetimes: Still subtle, Phys. Rev. D 98, 104007 (2018).
\bibitem{1001}
M. Rahman, S. Chakraborty, S. S. Guptab and A. A. Sen, Fate of strong cosmic censorship conjecture in presence of higher spacetime dimensions, JHEP 03, 178 (2019).

\end{thebibliography}
\end{document}